\documentclass{pnastwo}
\usepackage{amssymb,amsfonts,amsmath}
\usepackage{graphicx}
\usepackage{pnastwoF}
\usepackage{bm}

\contributor{Published in Proceedings 
of the National Academy of Sciences of the United States of America}
\url{www.pnas.org/cgi/doi/10.1073/pnas.1221976110}
\copyrightyear{2013}
\issuedate{Issue Date}
\volume{Volume}
\issuenumber{Issue Number}

\begin{document}


\title{Anomalous superfluid density in quantum critical superconductors} 


\author{K. Hashimoto\affil{1}{Department of Physics, Kyoto University, Kyoto 606-8502, Japan}\affil{5}{Present address: Institute for Materials Research, Tohoku University, Sendai 980-8577, Japan}, 
Y. Mizukami\affil{1}{}, R. Katsumata\affil{1}{},
H. Shishido\affil{1}{}\affil{6}{Present address: Department of Physics and Electronics, Osaka Prefecture University, Sakai, Osaka 599-8531, Japan}, 
M. Yamashita\affil{1}{}\affil{7}{Present address: RIKEN, Wako, Saitama 351-0198, Japan},
H. Ikeda\affil{1}{}, Y. Matsuda\affil{1}{}, \\
J.\,A. Schlueter\affil{2}{Materials Science Division, Argonne National Laboratory, Argonne, Illinois 60439, U.S.A.}, 
J.\,D. Fletcher\affil{3}{H. H. Wills Physics Laboratory, University of Bristol, Tyndall Avenue, Bristol, BS8 1TL, United Kingdom}, A. Carrington\affil{3}{}, 
D. Gnida\affil{4}{Institute of Low Temperature and Structure Research, Polish Academy of Sciences, P. O. Box 1410, 50-950 Wroc{\l}aw, Poland}, D. Kaczorowski\affil{4}{},
\and
T. Shibauchi\affil{1}{}\affil{8}{To whom correspondence should be addressed. Email: {\sf \small shibauchi@scphys.kyoto-u.ac.jp}}}

\contributor{Submitted to Proceedings of the National Academy of Sciences of the United States of America}

\maketitle

\begin{article}
\begin{abstract}
When a second-order magnetic phase transition is tuned to zero temperature by a non-thermal parameter, quantum fluctuations are critically enhanced, often leading to the emergence of unconventional superconductivity. In these `quantum critical' superconductors it has been widely reported that the normal-state properties above the superconducting transition temperature $T_c$ often exhibit anomalous non-Fermi liquid behaviors and enhanced electron correlations.  However, the effect of these strong critical fluctuations on the superconducting condensate below $T_c$ is less well established.   Here we report measurements of the magnetic penetration depth in heavy-fermion, iron-pnictide, and organic superconductors located close to antiferromagnetic quantum critical points showing that the superfluid density in these nodal superconductors universally exhibit, unlike the expected $T$-linear dependence, an anomalous 3/2 power-law temperature dependence over a wide temperature range.  We propose that this non-integer power-law can be explained if a strong renormalization of effective Fermi velocity due to quantum fluctuations occurs only for momenta $\bm{k}$ close to the nodes in the superconducting energy gap $\Delta(\bm{k})$. We suggest that such `nodal criticality' may have an impact on low-energy properties of quantum critical superconductors.
\end{abstract}
\keywords{superfluid stiffness | $d$-wave superconductivity | spin fluctuations | mass enhancement | quasiparticle scattering}


\dropcap{T}he physics of materials located close to a quantum critical point (QCP) 
is an important issue because the critical fluctuations associated with this point may produce unconventional high temperature superconductivity \cite{Varma02,Gegenwart08}.  Quantum oscillations \cite{Shishido05,Shishido10} and specific heat measurements \cite{Stewart01} have shown that, in some systems, as the material is tuned towards the QCP by controlling an external parameter such as doping, pressure, or magnetic field, the effective mass strongly increases due to enhanced correlation effects. Along with this the temperature dependence of the resistivity shows a strong deviation from the standard $AT^2$ dependence in the Fermi liquid (FL) theory of metals, and often shows an anomalous $T$-linear behavior which corresponds to the $A$ coefficient diverging as zero temperature is approached.

Although there are many studies of non-FL behavior in the normal metallic state \cite{Varma02,Gegenwart08}, relatively little is known about how the QCP affects the superconducting properties below the critical temperature $T_c$. The superconducting dome often develops around the putative QCP so that when the temperature is lowered below $T_c$, the superconducting order parameter starts to develop and the Fermi surface becomes gapped. It is therefore natural to consider that the low-energy quantum critical fluctuations are quenched by the formation of the superconducting gap $\Delta$, which means that the system avoids the anomalous singularities associated with the QCP.  Perhaps because of this reasoning the superconducting properties are usually analyzed by the conventional theory without including temperature/field dependent renormalization effects resulting from the proximity to the QCP.  For example, in Refs.\:\cite{Kohori01,Nakai10} the NMR relaxation rate $1/T_1$ in the superconducting state is fitted to the temperature dependence expected for particular gap functions with the assumption that the normal-state $1/T_1T$ is virtually temperature independent below $T_c$ even when it has strong temperature dependence above $T_c$ due to the magnetic fluctuations.

In superconductors near the QCP, the electron pairing is often unconventional with a superconducting energy gap $\Delta(\bm{k})$ which changes sign on different parts of the Fermi surface \cite{Sigrist91,Hirschfeld11}. This sign change stems from a repulsive pairing interaction for example resulting from antiferromagnetic spin fluctuations. In many cases, this leads to the presence of nodes in the gap $\Delta(\bm{k})$ where the gap changes sign. The low-energy excitations from the ground state in these superconductors are governed by these nodal regions. In such nodal superconductors, the effect of quantum critical fluctuations on the excited quasiparticles should be $\bm{k}$ dependent. Then the question arises as to how this effect modifies the low-energy properties in the superconducting state. 

The penetration depth $\lambda(T)$ is a fundamental property of the superconducting state which parameterizes the ability of a superconductor to screen an applied field by the diamagnetic response of the superconducting condensate. As fermionic quasiparticles are thermally excited from the condensate a paramagnetic current is created which reduces the screening and increases $\lambda$.  So measurements of $\lambda(T)$ give direct information about density and Fermi velocity of these quasiparticles \cite{ChandrasekharE93}. When the effective mass is enhanced by the quantum critical fluctuations, the effective Fermi velocity is expected to be suppressed accordingly. In a one-component Galilean invariant superfluid, electron correlation effects may not cause the renormalization in the low-temperature penetration depth \cite{Leggett65}. In superconducting materials, however, strong electron correlations do affect the renormalization resulting in an enhanced penetration depth, which has been reported both theoretically \cite{Varma86,Jujo02} and experimentally \cite{Gross86,Hashimoto12}. To discuss the energy-dependent effect of quantum criticality on superconducting quasiparticles, the temperature dependence of penetration depth at low temperatures is thus of particular importance. 

\section{Results}
\subsection{Penetration Depth}

Here we begin by presenting results for the heavy-fermion system Ce$_n$$T$In$_{3n+2}$ which is located close to a QCP ($T$ is the transition metal element, $n$ is the number of CeIn$_3$ layers alternating with the $T$In$_2$ blocks along the $c$ axis). The most studied member of this series is the $n = 1$ member CeCoIn$_5$ ($T_c = 2.3$\,K) \cite{Petrovic01}, in which clear evidence for non-FL behaviors in the normal state \cite{Nakajima07,Paglione03} and nodal superconductivity has been found \cite{Kohori01,Izawa01,An10}. The recent discovery of superconductivity at ambient pressure in the $n=2$ member Ce$_2$PdIn$_8$ ($T_c = 0.68$\,K) \cite{Kaczorowski09A}, which exhibits very similar non-FL properties \cite{Kaczorowski10,Dong11,Tokiwa11,Matusiak11,Gnida12,Fukazawa12} to those in CeCoIn$_5$, allows a detailed comparison of the superconducting properties to extract common features in these superconductors near the antiferromagnetic QCP. The magnetic penetration depth measurements have been made for high-quality single crystals of Ce$_2$PdIn$_8$ and CeCoIn$_5$ down to $\sim 60$\,mK by using a tunnel diode oscillator with the resonant frequency of $\sim 13$\,MHz (see {\it SI Text:\,SI1}). The weak ac field is applied along the $c$ axis, which excites supercurrents in the $ab$ plane.

\begin{figure*}[b]
\begin{center}
\includegraphics[width=1.0\linewidth]{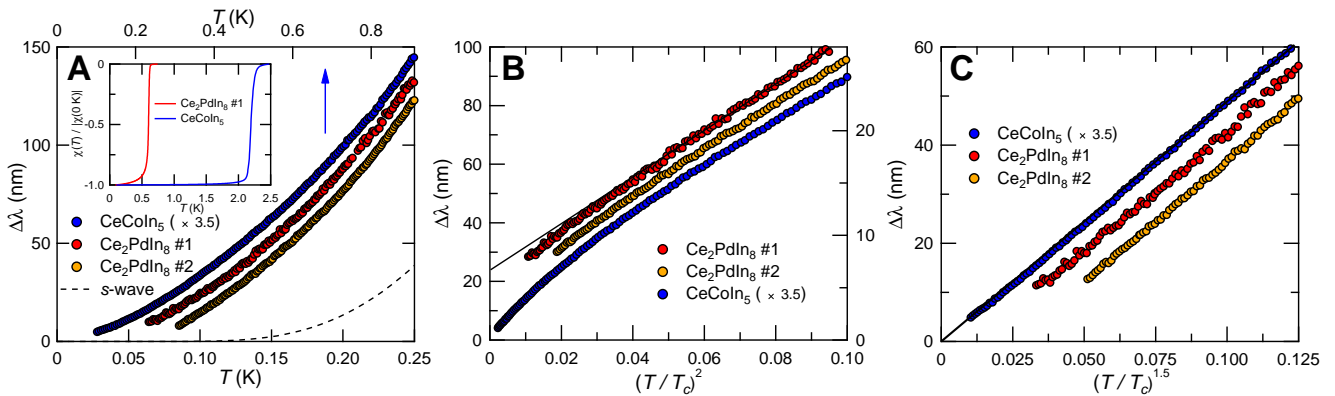}
\end{center}
\caption{Temperature dependence of the magnetic penetration depth in heavy-fermion superconductors near the antiferromagnetic QCP. ({\it A}): Low-temperature changes in the magnetic penetration depth $\Delta\lambda(T)=\lambda(T)-\lambda(T=0)$ of single crystals of Ce$_2$PdIn$_8$ and CeCoIn$_5$. The curves are shifted vertically for clarity and the data for CeCoIn$_5$ are multiplied by 3.5. The inset shows the ac susceptibility over the whole temperature range measured by the frequency shift of the tunnel diode oscillator, showing sharp superconducting transitions. The dashed line is an exponential temperature dependence expected for a fully gapped $s$-wave superconductor.
({\it B}): The same data plotted against $(T/T_c)^2$. The solid line represents a $T^2$ dependence.
({\it C}): The same data plotted against $(T/T_c)^{3/2}$. The solid line represents a $T^{3/2}$ dependence.}
\end{figure*}

The temperature dependent change in the in-plane penetration depth $\Delta\lambda(T)=\lambda(T)-\lambda(T=0)$ in both Ce$_2$PdIn$_8$ and CeCoIn$_5$ (Fig.\:1{\it A}) exhibits strong temperature variation at low-temperatures, much steeper than the flat exponential dependence expected for a fully gapped superconductor. The $\Delta\lambda(T)$ data for Ce$_2$PdIn$_8$ are reproducible in different crystals and the data for CeCoIn$_5$ are fully consistent with the previous studies \cite{Ormeno02,Chia03,Ozcan03}. The strong temperature dependence indicates substantial excitations of quasiparticles at low energies, evidencing the presence of line nodes in the energy gap. This is consistent with the residual density of states (DOS) in the low temperature limit observed by thermal conductivity \cite{Izawa01,Dong11}, specific heat \cite{An10,Tokiwa11}, and NMR measurements \cite{Kohori01,Fukazawa12}. In particular, a $d_{x^2-y^2}$ order parameter with nodes along the $\langle 1 1 0 \rangle$ directions has been established in CeCoIn$_5$ by angle-resolved thermal conductivity \cite{Izawa01} and specific heat measurements \cite{An10}. The strong similarity between Ce$_2$PdIn$_8$ and CeCoIn$_5$ found in the low-temperature $\lambda(T)$ points to common nodal structure in these superconductors.

In a pure $d$-wave superconductor with line nodes such as high-$T_c$ cuprates, it is well-established that $\Delta\lambda(T)$ shows a linear $T$-dependence at low temperatures, which stems from the linear energy dependence of low-energy DOS of quasiparticles. However, a clear deviation from the $T$-linear dependence is observed in the present heavy-fermion superconductors (Fig.\:1{\it A}). The data also strongly deviate from the $T^2$ dependence expected for the dirty limit case \cite{Hirschfeld93} (Fig.\:1{\it B}). We rather find that the power-law dependence $T^{\alpha}$ with an unusual exponent $\alpha=3/2$ can describe the observed low-temperature variation in a wide range in both superconductors (Fig.\:1{\it C}).

A few explanations for the super-linear temperature dependence of $\Delta\lambda(T)$ in $d$-wave superconductors have been put forth, including the effect of impurity scattering \cite{Hirschfeld93}, non-local effect near the nodes \cite{Kosztin97} and phase fluctuations \cite{Chen00}. Often an interpolation formula describing a crossover from $T$ to $T^2$ dependence $\Delta \lambda(T) \propto T^2/(T + T^*)$ is used to describe the experimental data. In particular, the impurity effect leads to the crossover temperature $T^*\approx 0.83\sqrt{\Gamma \Delta_0}$, where $\Gamma$ is the impurity scattering rate and $\Delta_0$ is the maximum gap. This successfully accounts for the systematic change of $\Delta\lambda(T)$ with impurity scattering observed in Zn-doped YBa$_2$Cu$_{3}$O$_{6.95}$ (Ref.\:\cite{Bonn94}). In the present heavy-fermion case, however, fitting to this crossover formula in these relatively clean superconductors yields $T^*/T_c$ values ($\sim 0.5$ for Ce$_2$PdIn$_8$ and $\sim 0.2$ for CeCoIn$_5$) substantially larger than the estimates from these theories (see {\it SI Text:\,SI2} and Ref.\:\cite{Ozcan03}). Therefore, the $T^{3/2}$ dependence in a wide $T/T_c$ range commonly observed in these superconductors with quite different $T_c$, which is distinctly different from the $T$-linear dependence in e.g. YBa$_2$Cu$_{3}$O$_{6.95}$ (Ref.\:\cite{Bonn94}), rather suggests some inherent mechanism related to their closeness to the antiferromagnetic QCP.

\subsection{Superfluid Density}

To discuss the precise temperature evolution of quasiparticle excitations, we analyze our data in terms of the superfluid density $1/\lambda^2(T)$. We use the reported values of $\lambda(0)$ (280\,nm for CeCoIn$_5$ \cite{Ozcan03} and 1010\,nm for Ce$_2$PdIn$_8$ \cite{Tran12}), from which the normalized superfluid density $\rho_s(T)=\lambda^2(0)/\lambda^2(T)$ has been obtained (Fig.\:2{\it A}). The factor of $\sim 3.5$ difference in the slope of $d\lambda/d(T/T_c)$ (Fig.\:1{\it A}) is consistent with the difference in $\lambda(0)$ in these two superconductors, which is also consistent with the larger $\gamma$ value in Ce$_2$PdIn$_8$ \cite{Tokiwa11}. This results in an almost collapse of the full temperature dependence $\rho_s(T)$ into a single curve (Fig.\:2{\it A}, inset), and the low-temperature variation shows the $(T/T_c)^{3/2}$ dependence with nearly identical slopes.

\begin{figure*}[t]
\begin{center}
\includegraphics[width=0.85\linewidth]{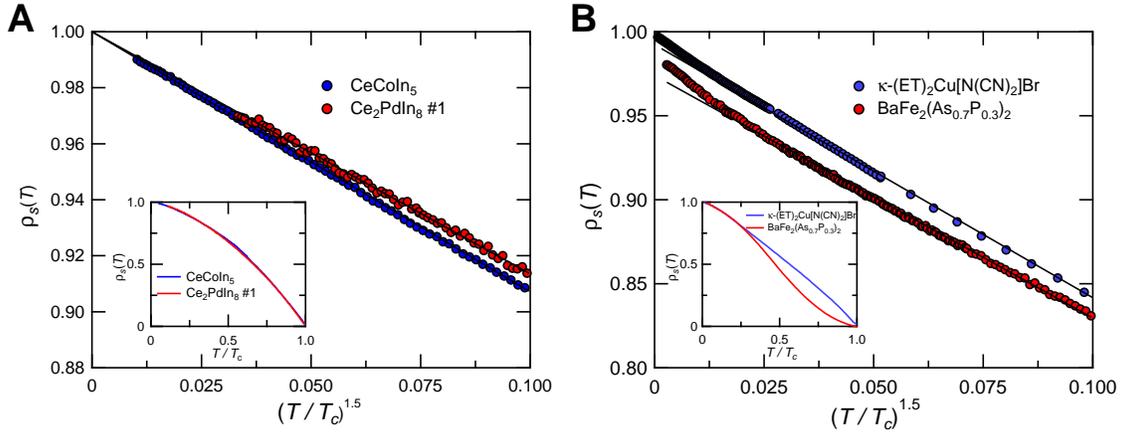}
\end{center}
\caption{Universal $T^{3/2}$ dependence of superfluid density in unconventional superconductors in the vicinity of the antiferromagnetic order. ({\it A}): The normalized superfluid density $\rho_s$ as a function of $(T/T_c)^{3/2}$ at low temperatures for Ce$_2$PdIn$_8$ and CeCoIn$_5$. The lines represent $T^{3/2}$ dependence. The inset shows the overall temperature dependence up to $(T/T_c)=1$. ({\it B}): A similar plot for iron-pnictide superconductor BaFe$_2$(As$_{0.7}$P$_{0.3}$)$_2$ and organic superconductor $\kappa$-(ET)$_2$Cu[N(CN)$_2$]Br. The solid lines are the fits to $T^{3/2}$ dependence. The low-temperature data for BaFe$_2$(As$_{0.7}$P$_{0.3}$)$_2$ are vertically shifted for clarity. 
} 
\end{figure*}

This $(T/T_c)^{3/2}$ dependence can be also seen in other classes of materials which are close to antiferromagnetic order.  In Fig.\:2{\it B} we show data for the organic superconductor $\kappa$-(BEDT-TTF)$_2$Cu[N(CN)$_2$]Br which is consistent with but measured to lower temperature than that in a previous report \cite{Carrington99} and the iron-pnictide superconductor BaFe$_2$(As$_{0.7}$P$_{0.3}$)$_2$ \cite{Hashimoto12}. Both these materials are known to have line nodes in their superconducting gap \cite{Hashimoto10,Taylor07}. In the BaFe$_2$(As$_{1-x}$P$_{x}$)$_2$ series, there is clear evidence for the antiferromagnetic QCP being located at $x\approx0.30$ \cite{Hashimoto12}. In $\kappa$-(ET)$_2$Cu[N(CN)$_2$]Br, although the proposed phase diagram suggests that the boundary between the superconducting and antiferromagnetic states is a first-order phase transition \cite{Kanoda06}, the anomalous critical exponent near the Mott critical end point \cite{Kagawa09} suggests the presence of strong antiferromagnetic quantum fluctuations. The normalized superfluid density $\rho_s(T)$ in these superconductors shows very similar $(T/T_c)^{3/2}$ dependence at low temperatures 
with slight deviations at the lowest temperatures below $T/T_c\sim 0.05$ ($(T/T_c)^{3/2}\lesssim 0.01$). 

These results imply that the $(T/T_c)^{3/2}$ dependence of superfluid density in a wide $T/T_c$ range is a robust property in unconventional superconductors, in which strong antiferromagnetic fluctuations are present (for comparisons between iron-pnictides and cuprates, see {\it SI Text:\,SI3}).

\section{Discussion}

If the quantum fluctuations survive the Fermi-surface gapping, the effective mass is expected to diverge in the zero temperature limit. In such a case, the strong enhancement of mass leads to a reduction of superfluid density when approaching the zero-temperature limit. It has been recently found in iron-pnictides that the zero-temperature superfluid density $1/\lambda^2(0)$ shows a strong reduction at the QCP \cite{Hashimoto12}, indicating the strong quantum critical fluctuations directly affect the superconducting condensate. However, the temperature dependence of the normalized superfluid density $\rho_s(T)$ actually continues to rise with decreasing temperature (Figs.\:2{\it A} and {\it B}, insets). This suggests that the effect of temperature dependent quantum fluctuations enters as corrections in the temperature dependence of superfluid density at low temperatures. The anomalous non-integer power-law temperature dependence of $\rho_s(T)$ universally observed in quantum critical superconductors thus calls for further theoretical understanding.

\subsection{Nodal Quantum Criticality}

Below we will propose a possible scenario that this universal behavior can be naturally explained by invoking a strong momentum dependence of renormalization due to the nodal gap structure. Suppose that the gap formation below $T_c$ quenches the low-energy quantum critical fluctuations, which prevents the effective mass enhancement in the superconducting state. Then in the nodal superconductors, because of the strong momentum dependence of the gap magnitude we may consider that the degree of quenching of the quantum fluctuations in the superconducting state has a strong momentum dependence as well; the low-energy fluctuations are expected to be strongest near the nodes where the Fermi surface is not gapped (Fig.\:3{\it A}). To model this effect, we consider the angle dependence of the effective Fermi velocity $v^*_F(\phi)$ along the underlying Fermi surface. We consider a simple cylindrical Fermi surface and assume that the renormalization in $v^*_F(\bm{k})/v_F$ is inversely related to the enhancement in the effective mass $m^*(\bm{k})$ (Fig.\:3{\it B}). (Here $1/v^*_F$ is given by the dynamic effective mass which is different from, but closely related to the thermodynamic mass \cite{Varma86}). The effective mass enhancement on approaching the QCP can be described by $m^{*2} \propto (p-p_{\rm QCP})^{-\beta}$, where $p$ is a nonthermal parameter controlling the distance from the QCP at $p_{\rm QCP}$. The critical exponent $\beta$ has been estimated experimentally by using magnetic fields as the parameter $p$, and in several materials a value of $\beta$ close to unity has been reported \cite{Gegenwart02,Matsumoto11}. In the present case, we take the magnitude of the superconducting gap $|\Delta|$ as the control parameter, because the gap magnitude determines the degree of quenching of low-energy fluctuations. We thus assume $v^*_F(\bm{k})\propto |\Delta(\bm{k})|^{\beta/2}$, which allows us to calculate the temperature dependence of superfluid density by the integral over the Fermi surface $\bm{S}$ as \cite{ChandrasekharE93,Jujo02}
\begin{equation}\label{Eq:lam}
\lambda_{ij}^{-2}(T)=
\frac{\mu_0e^2}{4\pi^3\hbar}\int \frac{v_{Fi}(\bm{k})v^*_{Fj}(\bm{k})}{|\bm{v}_{F}(\bm{k})|}
\left[1-Y(\bm{k},T)\right]
d\bm{S},
\end{equation}
where $v^*_F$ ($v_F$) is the effective velocity in the superconducting (normal) state, the subscripts $i,j$ denote the directions of the current and vector potential (we take both along $a$), $Y(\bm{k},T)=
-2 \int_{\Delta(\bm{k})}^\infty \frac{\partial f(E)}{\partial E}\frac{E}{\sqrt{E^2-\Delta^2(\bm{k})}} dE$ is Yosida function, and $f(E)$ is the Fermi-Dirac function for the quasiparticle energy $E$. By using the $d_{x^2-y^2}$ formula of $\Delta(\bm{k})=\Delta_0 \cos(2\phi)$ and $\beta=1$, the normalized superfluid density $\rho_s$ is calculated with $v^*_F(\phi)\propto |\cos(2\phi)|^{1/2}$ (Fig.\:3{\it C}), which results in the $T^{3/2}$ dependence at low temperatures (Fig.\:3{\it D}). 

As real materials will never be located exactly at the QCP there will be a cutoff for the diverging $m^*$ near the nodes. This can be modelled by introducing a minimum value for $v^*_F$. This leads to an upward deviation from the $T^{3/2}$ dependence of $\rho_s$ at very low temperatures approaching $T$-linear behavior at sufficiently low temperatures (Fig.\:3{\it D}). This can explain the essential features of the experimental observations. We note that the presence of the cutoff can be also expected even at the exact QCP, because the dynamical susceptibility in the zero temperature limit should diverge only on certain regions of the Fermi surface (determined by the momentum dependence of the spin fluctuations), which in general may be different from the nodal points \cite{Hirschfeld11}. Of course, disorder would also produce some additional changes to the temperature dependence but we have not included this in the present model.

\begin{figure*}[t]
\begin{center}
\includegraphics[width=0.9\linewidth]{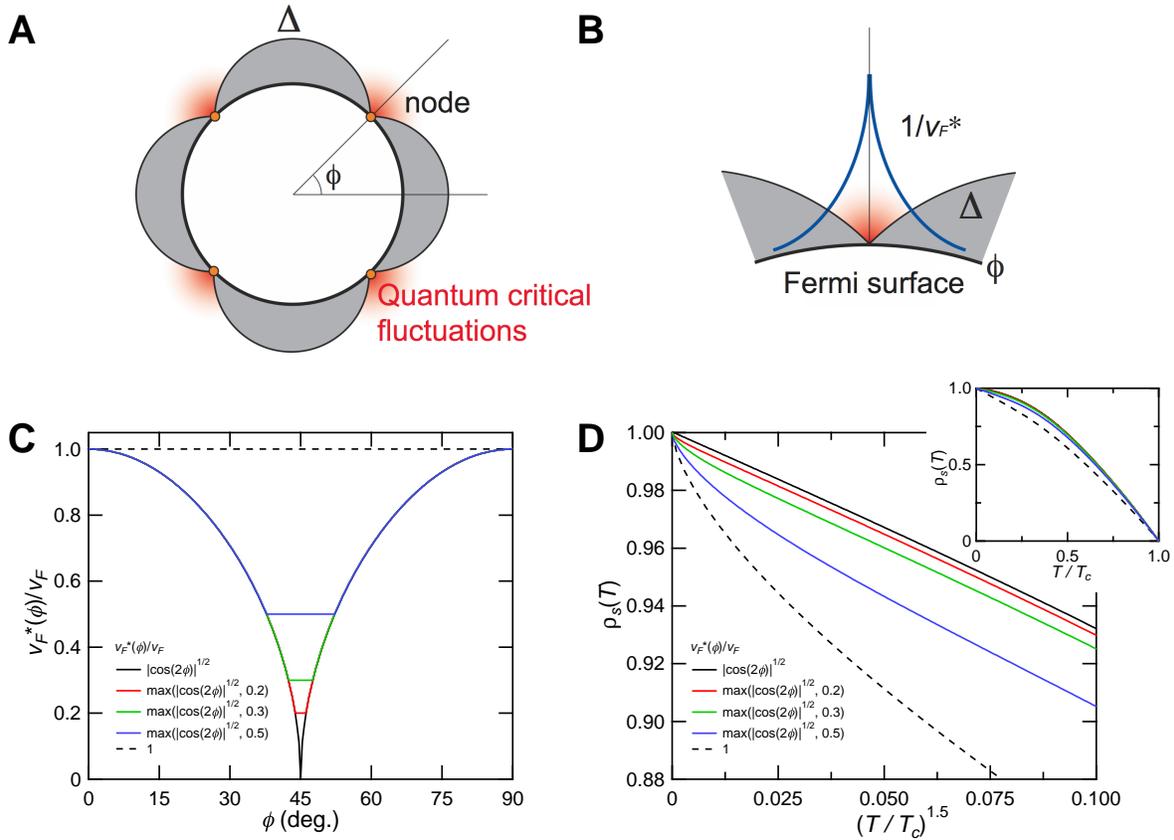}
\end{center}
\caption{Nodal quantum criticality in unconventional superconductors. ({\it A}): The momentum dependent gap $\Delta(\bm{k})$ (whose magnitude is illustrated by thin lines with grey shades) opens on the Fermi surface (thick line), and has nodes (red circles) at certain directions. In $d_{x^2-y^2}$-wave superconductors, for example, $\Delta(\bm{k})$ has strong in-plane anisotropy $\Delta_0 \cos(2\phi)$ as a function of azimuthal angle $\phi$. In quasi-two dimensional systems, the Fermi surface is approximated by a cylinder, and thus the gap has nodal lines at perpendicular to the planes. At the nodes, the gap is zero and thus the quantum critical fluctuations may be present (red shades) on the ungapped Fermi surface. ({\it B}): The nodal quantum fluctuations lead to the momentum dependence of the renormalization in $1/v^*_F(\bm{k})$ near the nodes (blue lines). 
({\it C}): The angle dependence of the renormalized Fermi velocity $v^*_F(\phi)$ relative to the unrenormalized one $v_F$ along the Fermi surface, assumed for calculations of the superfluid density in ({\it D}). Near the nodes, we illustrate different cutoff levels, which model finite distances from the QCP or disorder. ({\it D}): Calculated normalized superfluid density as a function of $(T/T_c)^{3/2}$ with different cutoff levels, which explains the deviation from the $T^{3/2}$ dependence at very low temperatures. The inset is the full temperature dependence up to $(T/T_c)=1$.}
\end{figure*}

An additional factor could also come from the temperature dependence of the renormalization of $m^*$.  As only the thermally excited quasiparticles will be renormalized, the angular range near the nodes where this occurs (i.e., where $\Delta(\bm{k})<k_BT$) is quite limited at low temperature and so this has a rather minor effect. In our temperature range of interest, $T\lesssim 0.2T_c$, this region is actually limited to a narrow angle range of $\sim\pm 3^\circ$ near the nodes. If we add a cutoff constant of 0.3 covering this angle range (Fig.\:3{\it C}), we found that the $T$-dependence is affected only in the lowest temperature range of $(T/T_c)^{1.5}\lesssim 0.01$, above which the $T^{3/2}$ dependence of $\rho_s$ still holds (Fig.\:3{\it D}). This exercise implies that the inclusion of the temperature dependence of renormalization will not change $\rho_s(T)$ significantly in the temperature range of interest (for more discussion, see {\it SI Text:\,SI4}). 

We also note that in iron-pnictides the gap symmetry is most likely $s$-wave \cite{Yamashita11}, and the model based on the $d$-wave gap may not be applicable. However, the fundamental physics that the low-energy excitations are governed by the nodal regions should be essentially the same. Although the detailed structure of the momentum-dependent Fermi velocity (such as the precise value of the critical exponent $\beta$) will affect the detailed $\rho_s(T)$ (see {\it SI Text:\,SI5}), it is striking that such a simple model can capture the salient feature of the unusual behavior of quasiparticle excitation in the superconducting state of these materials. In a FL theory, the renormalization of the effective Fermi velocity $v^*_F$ in Eq.\:[1] can be described by the interplay between $\bm{k}$-dependent quasiparticle interaction and nodal gap structure \cite{Jujo02}. More detailed theoretical calculations will be needed to account for: strong energy dependence of critical fluctuations, the effects of disorder (such as quasiparticle scattering interference \cite{Lee03}) and the inelastic quasiparticle scattering rate. 

We suggest that the nodal quantum criticality proposed here is an important aspect of unconventional superconductivity close to the magnetic QCP. Besides the penetration depth analyzed here there will be implications for most other superconducting properties such as thermal conductivity and the NMR relaxation rate $1/T_1(T)$ which has been long known to have strong deviations from the $T^3$ law, which in the usual analysis gives overestimates of the residual quasiparticle DOS in quantum critical superconductors \cite{Kohori01,Nakai10}.  
It should be straightforward to extend our analysis to these other properties.

\begin{acknowledgments}
We thank A. Chubukov, H. Fukazawa, R.\,W. Giannetta, K. Kanoda, S. Kasahara, H. Kontani, S.\,K. Goh, P.\,J. Hirschfeld, A.\,H. Nevidomskyy, T. Nomoto, R. Prozorov, I. Vekhter, Y. Yanase, and Y.\,F. Yang for discussions. This work is supported by KAKENHI from JSPS, Grant-in-Aid for GCOE program ``The Next Generation of Physics, Spun from Universality and Emergence'' from MEXT, Japan, EPSRC in the UK, the National Science Centre (Poland) under the research Grant No. 2011/01/B/ST3/04482 in Poland and by Argonne, a U.S. Department of Energy Office of Science laboratory, operated under Contract No. DE-AC02-06CH11357 in the USA.
\end{acknowledgments}

\end{article}



\clearpage





\renewcommand{\thefigure}{S\arabic{figure}}
\setcounter{figure}{0}

\begin{article}

\noindent
{\huge\bf Supporting Information} 

\bigskip
\noindent
{\large\bf Hashimoto {\it et al.} 10.1073/pnas.1221976110} 

\section{SI Text}

\section{SI1--Sample information and experimental techniques}

High-quality single crystals of Ce$_2$PdIn$_8$ and CeCoIn$_5$ were grown by the self-flux method \cite{SShishido02,SKaczorowski09A}. The typical lateral size of the crystals is $150\times150\,\mu$m$^2$. For Ce$_2$PdIn$_8$ very thin crystals with a thickness of less than 10 $\mu$m were prepared to avoid contamination of CeIn$_3$. We have checked that our crystals used in this study show no antiferromagnetic signals. The magnetic penetration depth measurements down to very low temperatures ($\sim 60$\,mK) have been performed by using a tunnel diode oscillator with the resonant frequency of 13\,MHz \cite{SCarrington99,SHashimoto10}, which is mounted on a $^3$He-$^4$He dilution refrigerator. The sample is placed on a sapphire rod, the other end of which is glued to a copper block on which a RuO$_2$ thermometer is mounted. The sample is placed at the center of a solenoid which forms part of the resonant tank circuit. A weak ac field is applied along the $c$-axis so that the supercurrent flows within the $ab$-plane. Changes in the resonant frequency are directly proportional to changes in the magnetic penetration depth, $\Delta\lambda(T)=G\Delta f(T)$. The calibration factor $G$ is determined from the geometry of the sample \cite{SProzorov00}.

Samples of $\kappa$-(BEDT-TTF)$_2$Cu[N(CN)$_2$]Br were grown via the standard electrocrystalisation method \cite{SKini90}. Measurements were also made using a tunnel diode oscillator. Special care was taken to cool the samples slowly across the temperature region 70 -- 90\,K where a structural ordering transition occurs. The cooling rate was restricted to 0.02\,K/min, which is sufficiently slow enough to avoid suppression in $T_c$ and reduction in electronic contribution to the heat capacity \cite{STaylor08}.

The data in Fig.\:2{\it B} was taken for a sample with dimensions $0.8\times 0.7\times 0.4$\,mm$^3$. In these organic samples even weak radio frequency (RF) fields can produce self-heating effects at the lowest temperatures ($T < 300$\,mK). To eliminate this effect, measurements were made with the sample at different positions on the axis of the RF coil, effectively reducing the applied field. Typically, a reduction in field of a factor three was sufficient to eliminate all signs of self-heating. This procedure was tested in three samples and consistent results were obtained.

\section{SI2--Comparisons with existing theories}

The origins of the deviation from the $T$-linear temperature dependence of $\lambda(T)$ in a $d$-wave superconductor are discussed in the framework of two different theories. Hirschfeld and Goldenfeld \cite{SHirschfeld93} have shown that a small amount of impurity causes a residual density of states near $E_F$, which can change the linear-$T$ dependence (pure regime) to a quadratic $T$-dependence (impurity-dominated regime) at a crossover temperature $T^{\ast}_{\rm imp}$. This characteristic temperature is given by $T^{\ast}_{\rm imp}\approx 0.83\sqrt{\Gamma \Delta_0}$, where $\Gamma$ is the impurity scattering rate and $\Delta_0$ is the maximum gap. On the other hand, Kosztin and Leggett \cite{SKosztin97} have pointed out that since the coherence length $\xi$ diverges near the nodes in a $d$-wave superconductor, Cooper pairs near the nodal directions, where $\xi \gg \lambda$, cannot participate in the field screening. Therefore, the effective penetration depth should be larger than the local limit value below a characteristic temperature $T^{\ast}_{NL}\approx \Delta_0 \xi(0)/\lambda(0)$, which leads to a quadratic $T$-dependence. Thus, for both theories the interpolation formula between two regions is given by $\Delta \lambda (T) \propto T^2/(T + T^{\ast})$.
Here we note an important distinction between the temperature dependence of $\Delta\lambda(T)$, which is directly measured, and the normalised superfluid density, $\rho_s(T) = \lambda^2(0)/\lambda^2(T)$. The $d$-wave form for the superfluid density, $\rho_s(T) = 1-\mu T/T_c$, gives $\Delta\lambda(T)/\lambda(0) = \frac{1}{2}\mu(T/T_c)+\frac{3}{8}\mu^2(T/T_c)^2+\cdot \cdot \cdot$, which lead to a small quadratic component to $\lambda(T)$ that depends on $\mu$, where $\mu$ is proportional to the inverse of the slope of the gap $\Delta(\phi)$ at the nodes, $\mu = (4\ln 2) k_BT_c/(d\Delta(\phi)/d\phi |_{\rm node})$ \cite{SXu95}. Then, even in the pure $d$-wave model we have slightly concave temperature dependence for $\lambda(T)$. Therefore, we take the following formula to describe the crossover behavior expected in these theories;
\begin{equation}
\rho_s(T) = 1-\mu (T^2/T_c)/(T+T^{\ast}). \tag{S-1}
\label{crossover}
\end{equation}

%
\begin{figure}[t]
\begin{center}
\includegraphics[width=\linewidth]{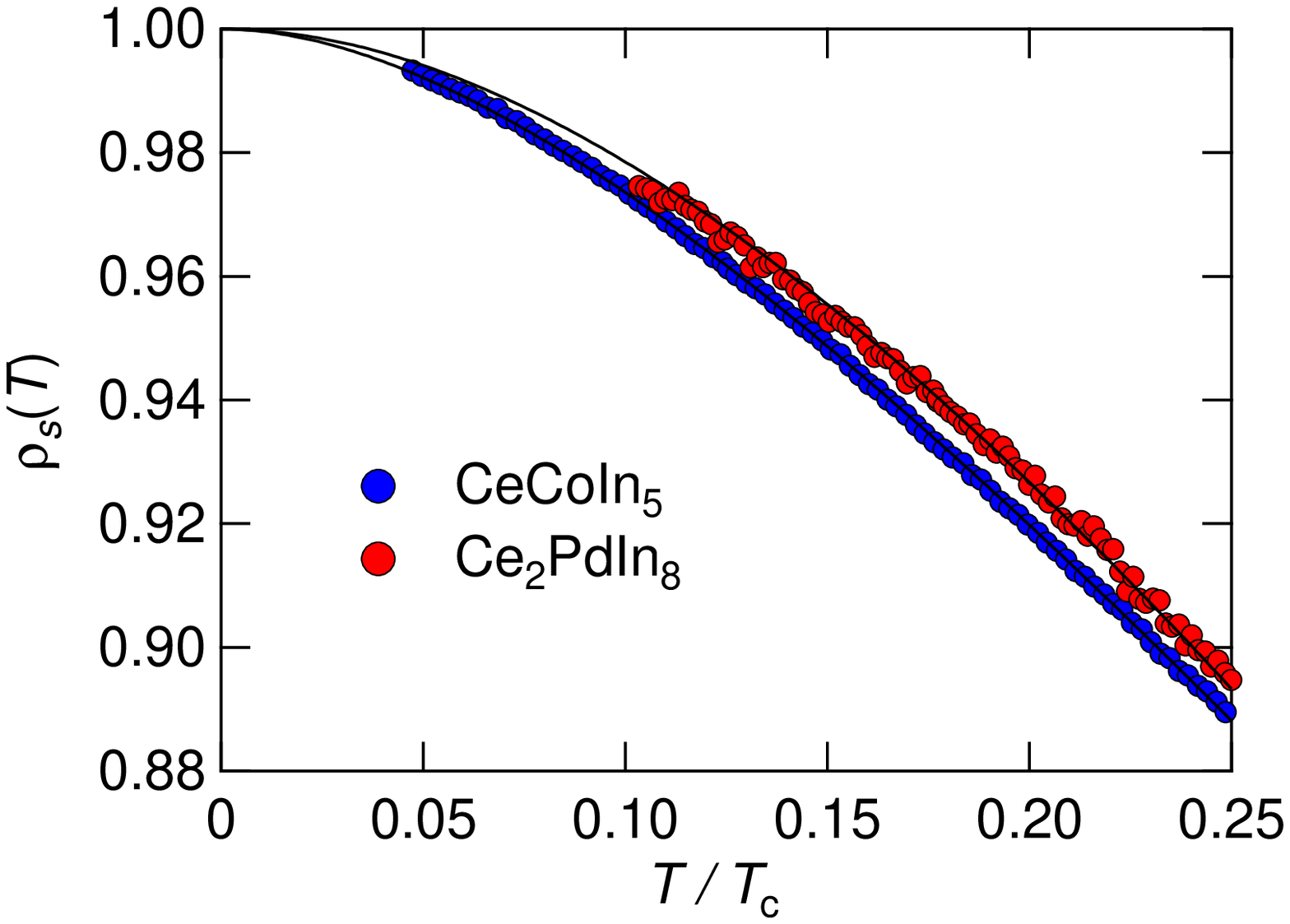}
\end{center}
\caption{
\setcounter{figure}{1}
Expanded view of normalized superfluid density plotted against $T/T_c$ for Ce$_2$PdIn$_8$ and CeCoIn$_5$. The solid lines are fit to the crossover behavior in a $d$-wave superconductor described in the text.
} 
\end{figure}

Figure\:S1 shows the low-temperature superfluid density $\rho_s(T)$ both for Ce$_2$PdIn$_8$ and CeCoIn$_5$. We carefully examine the possibility of the two theories mentioned above by evaluating $T^{\ast}_{\rm exp}$ from the present data. The obtained values of $T^{\ast}_{\rm exp}/T_c$ from the fits are 0.48 and 0.22 for Ce$_2$PdIn$_8$ and CeCoIn$_5$, respectively. If we assume that the origin of nonlinearity of $\lambda(T)$ comes from the residual density of states induced by impurities, we can estimate the impurity scattering rate $\Gamma$ via $T^{\ast}\approx 0.83\sqrt{\Gamma \Delta_0}$. The obtained value of $\Gamma$ for Ce$_2$PdIn$_8$ is 0.11 K. By using this and Abrikosov-Gor'kov theory\cite{SGor'kov85}, we estimated that the $T_c$ reduction is 11\% for Ce$_2$PdIn$_8$. However, recent specific heat measurements have suggested that the impurity level of Ce$_2$PdIn$_8$ is comparable to 0.22\% Cd-doped CeCoIn$_5$ \cite{STokiwa11}. In this case, $T_c$ reduction is less than 5 \%. Similarly, the obtained value of $T^{\ast}$ for CeCoIn$_5$ using Eq.\:[S-1] is much larger than the value estimated from thermal transport measurements ($<$ 30 mK) \cite{SMovshovich01}. Therefore, impurity effect are unlikely to explain the nonlinearity of $\lambda(T)$ in these systems.  As for the nonlocal effects, the reported values of $\xi_0 \sim 8$\,nm \cite{SKaczorowski09A,SDong11,SSettai01} for both compounds give $T^{\ast}_{NL} \approx$ 14\,mK and 140\,mK for Ce$_2$PdIn$_8$ and CeCoIn$_5$, respectively, which indicates that the non-locality effect should be important at much lower temperatures than our $T^{\ast}_{\rm exp}$. Thus, we can exclude these effects as primary origins of the deviation from the $T$-linear behavior in $\rho_s(T)$.

We also note that the $T^{3/2}$ dependence of low-temperature superfluid density has been derived from the Bardeen-Cooper-Schrieffer (BCS) to Bose-Einstein condensation (BEC) crossover below $T_c$ \cite{SChen00}. However, this assumes an $s$-wave state, which is not applicable to the nodal superconductors focused in the present study.

\section{SI3--Comparisons between iron-pnicteides and cuprates}

We comment on the difference in the high-$T_c$ cuprate superconductors. In clean single crystals of optimally doped YBa$_2$Cu$_3$O$_{7-\delta}$, the superfluid density exhibits a rather wide range of $T$-linear behavior and the crossover temperature is as low as $T^{\ast}_{\rm exp}/T_c\sim 0.01$. In hole-doped cuprates, the superconducting dome is separated from the antiferromagnetic order, and a possible QCP of the pseudogap phase has been discussed near the optimum doping of the superconducting dome. The absolute value of the superfluid density shows a broad maximum near the optimal doping \cite{STallon03}, which is opposite to the iron-pnictide BaFe$_2$(As$_{1-x}$P$_x$)$_2$ case, where the sharp minimum is observed at the antiferromagnetic QCP \cite{SHashimoto12}. The latter case is consistent with the enhanced mass (which enters inversely to the superfluid density) associated with the QCP. This difference in doping dependence seems consistent with the difference in the temperature dependence of superfluid density; the $T^{3/2}$ dependence is a common property associated with the mass enhancement due to the antiferromagnetic QCP.

To see the doping dependence of superfluid density in the iron-pnictide BaFe$_2$(As$_{1-x}$P$_x$)$_2$, we plot the low-temperature $\rho_s$ as a function of $(T/T_c)^{1.5}$ in a wide doping range (Fig.\:S2). The antiferromagnetic QCP composition in this system has been located at $x=0.30$ \cite{SHashimoto12}. In high composition samples ($x\geq0.38$) which are relatively far from the QCP, the low-temperature deviations from the $T^{3/2}$ dependence is more pronounced than those in samples closer to the QCP ($0.27\leq x\leq 0.33$). These results further support that the non-integer power law is related to the closeness to the magnetic QCP. In hole-doped cuprate superconductors, the antiferromagnetic QCP is actually located outside the low-doping end point of the superconducting dome. It deserves further studies to understand how the phase diagrams relate to these differences of low-energy quasiparticle excitations in cuprates and iron-based high-$T_c$ superconductors.

\begin{figure}[h]
\begin{center}
\includegraphics[width=\linewidth]{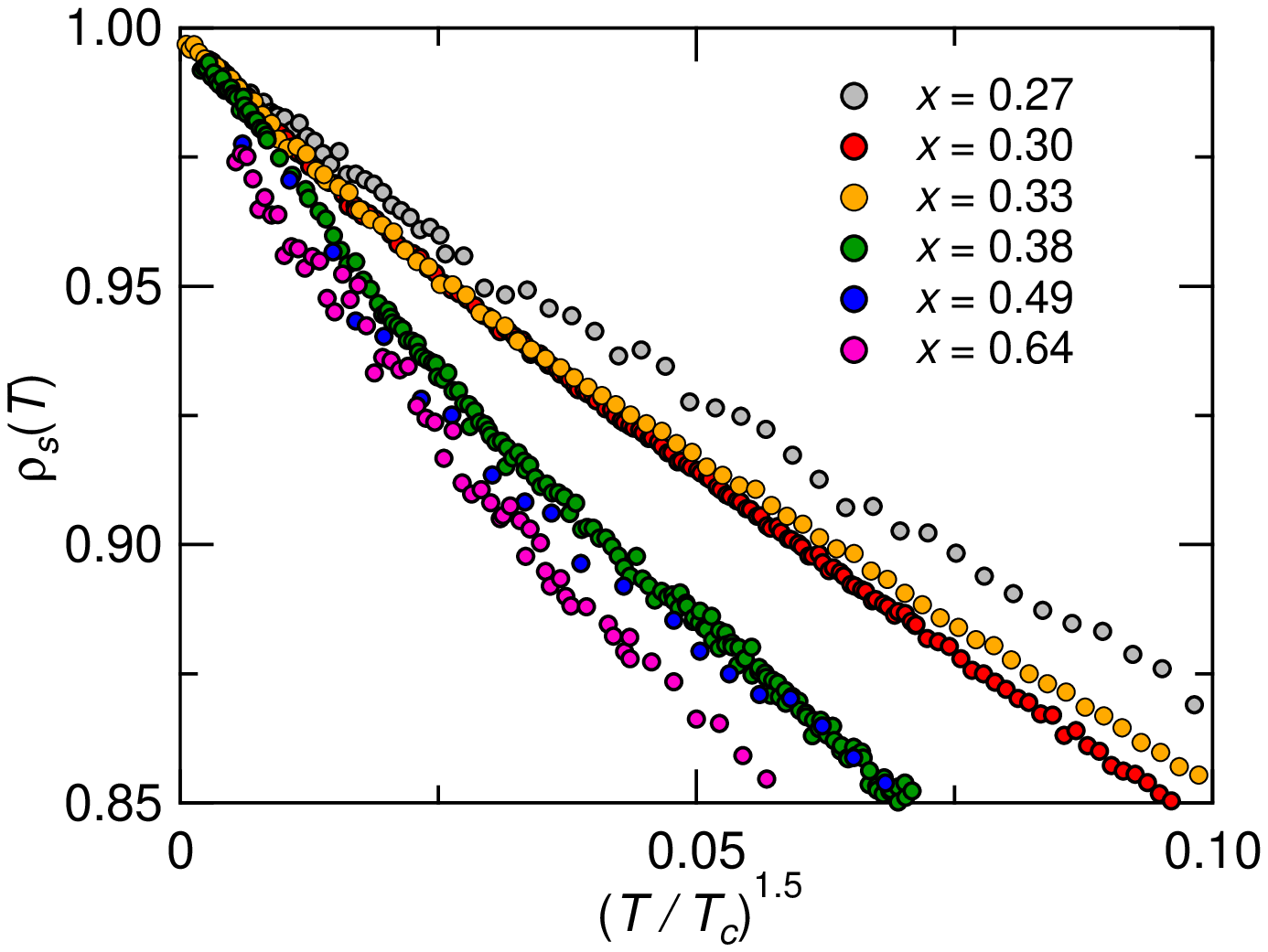}
\end{center}
\caption{
\setcounter{figure}{2}
The normalized superfluid density $\rho_s$ as a function of $(T/T_c)^{1.5}$ at low temperatures for BaFe$_2$(As$_{1-x}$P$_{x}$)$_2$ single crystals. The original penetration depth data are taken from Ref.\:\cite{SHashimoto12}.
} 
\end{figure}

\section{SI4--Temperature dependence of the renormalization}

Near the QCP, the temperature dependence of the renormalization is an important factor. In the normal state above the upper critical field, the specific heat is largely enhanced at low temperatures \cite{SBianchi03}, indicating the strong temperature dependence of $m^*$ which may be approximated by $1/\sqrt{T}$ at low temperatures (Fig.\:S3{\it A}). (Note that due to the field-induced QCP, this dependence is the strongest near the upper critical field and the actual temperature dependence at zero field may be weaker.) In the superconducting state, owing to the momentum dependence of the renormalization we consider, $1/v^*_F(T)$ will level off at momentum dependent temperature. Near the nodes, the temperature dependence of renormalization will continue to low temperatures, whereas in the antinodal directions it will become independent of temperature near $T_c$.

If we assume $v^*_F(\phi,T)/v_F=\max(C\sqrt{T/T_c},|\cos(2\phi)|^{1/2})$, we can calculate the temperature dependence of averaged mass enhancement in the superconducting state (Fig.\:S3{\it A}). Here we separate the two terms in Eq.\:[1] in the main text as $\lambda^{-2}=\lambda_d^{-2}-\lambda_p^{-2}$, where the first term 
$\lambda_d^{-2}=\frac{\mu_0e^2}{4\pi^3\hbar}\int \frac{v_{F}v^*_{F}}{|\bm{v_{F}}|}d\bm{S}$ corresponds to the diamagnetic current and the second term comes from the paramagnetic current carried by the thermally excited quasiparticles. In Fig.\:S3{\it B} the temperature dependence of each contribution is presented with and without the temperature dependence of renormalization. These results indicate that in the superconducting state the temperature dependence of the averaged renormalization is weak, which seems to be consistent with the recent high-frequency conductivity results \cite{SBroun12}. Moreover, the temperature dependence of the total superfluid density $\lambda^{-2}(T)$ remains almost unchanged even we consider the temperature dependence of the renormalization in the superconducting state. 

\begin{figure*}[b]
\begin{center}
\includegraphics[width=\linewidth]{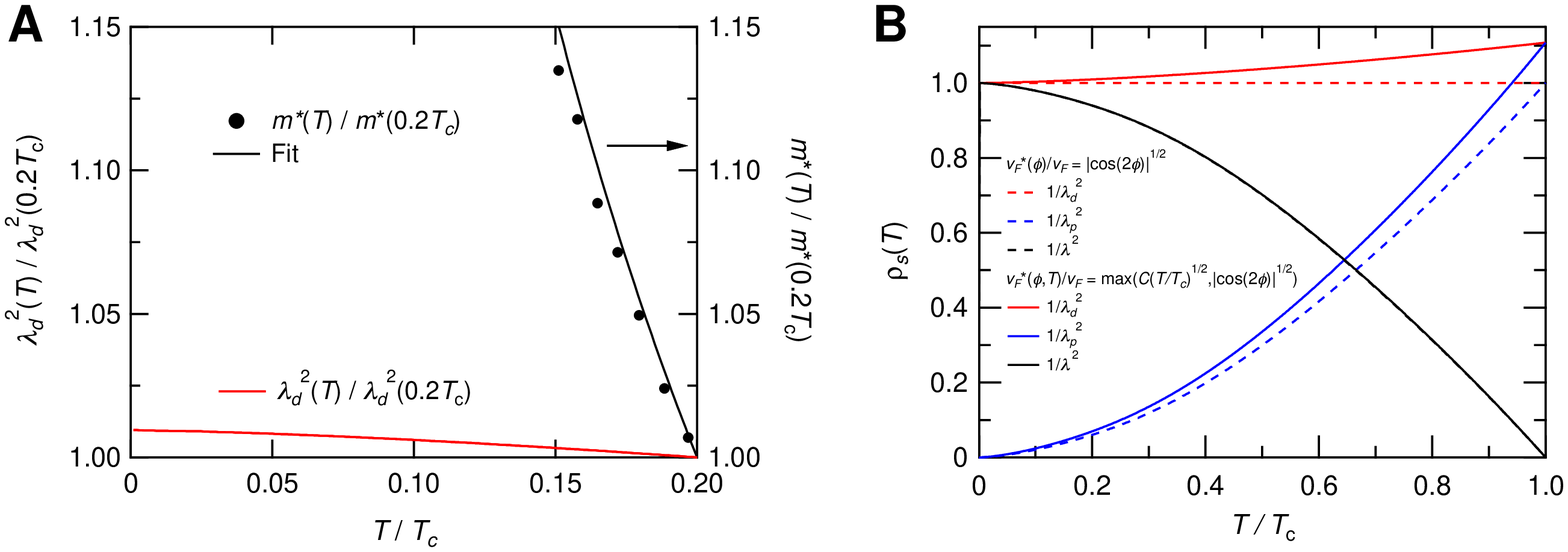}
\end{center}
\caption{
\setcounter{figure}{3}
({\it A}): Temperature dependence of the effective mass normalized by its $0.2T_c$ value in CeCoIn$_5$ (black circles) determined from the specific heat measured at 5\,T in the normal state (above the upper critical field) \cite{SBianchi03} can be fitted to a $1/\sqrt{T}$ dependence (black line). The inverse of the effective Fermi velocity averaged over the Fermi surface in the superconducting state (corresponding to the first term in Eq.\:[1] in the main text) calculated assuming $v^*_F(\phi,T)/v_F=\max(C\sqrt{T/T_c},|\cos(2\phi)|^{1/2})$ with a proper constant $C=\sqrt{5}|\cos2(45^\circ\pm3^\circ)|^{1/2}$, has a much weaker temperature dependence (red line). ({\it B}): Temperature dependence of the normalized superfluid density with (solid lines) and without (dashed lines) assuming the temperature dependence of renormalization. The total superfluid density (black) can be separated to the diamagnetic (red) and paramagnetic (blue) components. \\
} 
\end{figure*}

\section{SI5--Effect of quantum critical exponent}

To see how the power-law of superfluid density varies with the quantum critical exponent $\beta$, we have computed $\rho_s(T)$ with several values of $\beta$ up to $\sim 2$. The low-temperature $1-\rho_s(T)$ is fitted to the $T^\alpha$ dependence up to $T=0.2T_c$ and the obtained power $\alpha$ is plotted against the exponent $\beta$ in Fig.\:S4. We find that the power $\alpha$ varies linearly with $\beta$, which can be approximated by $\alpha\approx 1+\beta/2$.

The fact that we universally obtained $\alpha$ close to 3/2 in the present quantum critical superconductors suggests that the exponent $\beta$ in these materials is not far from unity. Indeed, in $\beta$-YbAlB$_4$ in which zero-field quantum criticality has been suggested \cite{SMatsumoto11}, thermodynamic measurements shows $\beta\approx 1$, which leads to $\alpha\approx 1.54$ (Fig.\:S4). In the study of YbRh$_2$Si$_2$, where the quantum critical field is 0.66\,T, $\beta\approx 1$ has also been reported \cite{SGegenwart02}. We note that in a similar system YbRh$_2$(Si$_{0.95}$Ge$_{0.05}$)$_2$ where the critical field is reduced to 0.027\,T, smaller value of $\beta\approx 2/3$ has been suggested \cite{SCusters03}, but this value gives $\alpha\approx1.36$, which is in practise difficult to distinguish from the 3/2 power-law dependence. We also note that in CeCoIn$_5$ and Ce$_2$PdIn$_8$ superconductors the estimated values for $\beta$ from transport coefficient $A$ near the upper critical field $H_{c2}$ are rather scattered ($\sim 1.37$ for CeCoIn$_5$ \cite{SPaglione03} and $\sim 0.57$ for Ce$_2$PdIn$_8$ \cite{SDong11}), but the first-order transition of the low-temperature $H_{c2}$ due to the strong Pauli paramagnetism \cite{SIzawa01,STokiwa11} may provide some complexity in evaluating the precise value of critical exponent.

\begin{figure}[t]
\begin{center}
\includegraphics[width=\linewidth]{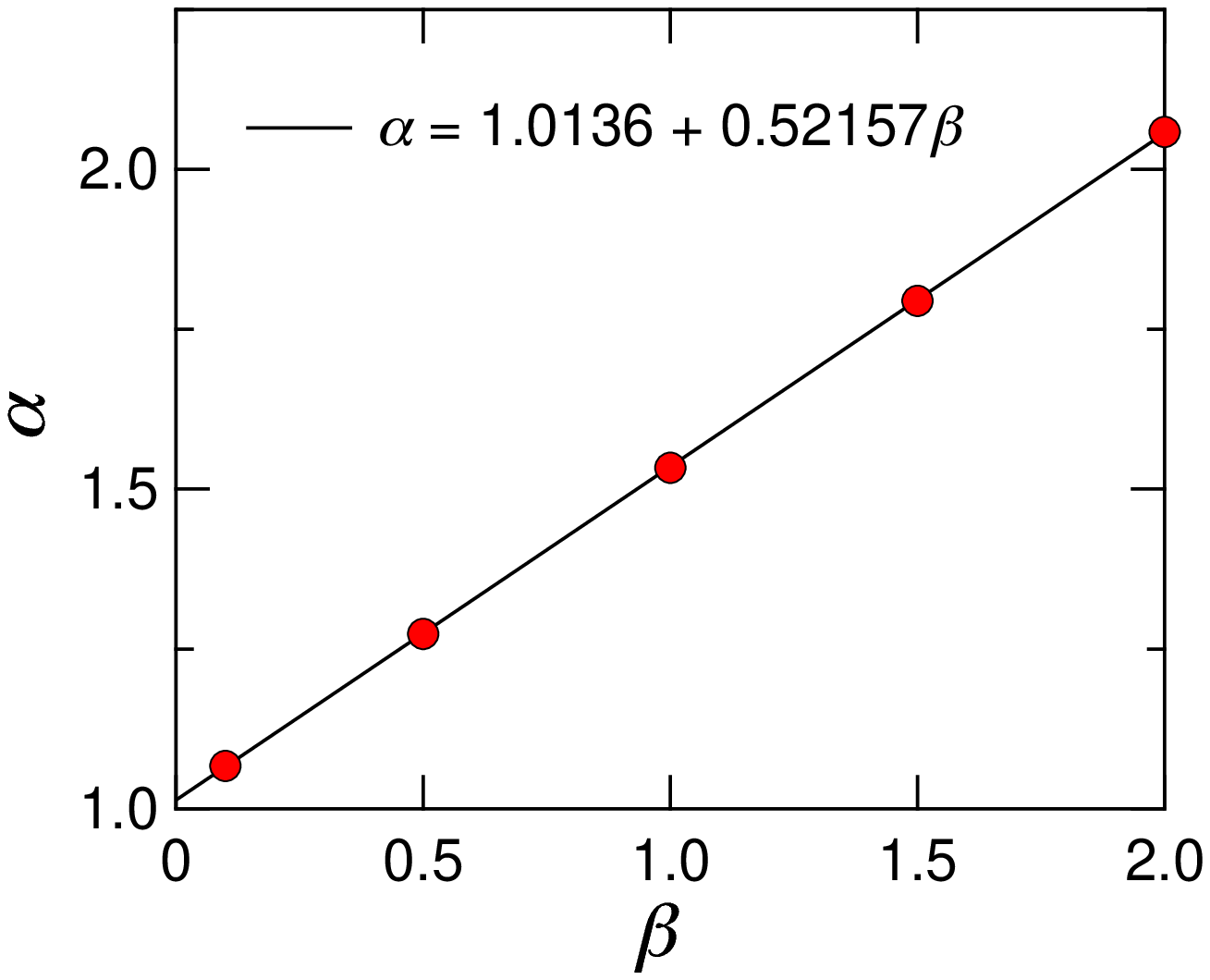}
\end{center}
\caption{
\setcounter{figure}{4}
Low-temperature power $\alpha$ of the superfluid density obtained from the fitting up to $T/T_c=0.2$ for the calculated $\rho_s(T/T_c)$ with different values of the quantum critical exponent $\beta$. The line is a fit to a linear dependence.
} 
\end{figure}




\end{article}
\end{document}